\documentclass{elsart}
\usepackage{amsfonts,amsmath,amssymb,epsfig,graphicx,algorithm,subfig}
\usepackage[usenames,dvipsnames]{xcolor}

\journal{Journal of Computational Physics}

\begin{document}
\begin{frontmatter}

\title{Forward and Adjoint Sensitivity Computation of Chaotic Dynamical Systems}
\author[mit]{Qiqi Wang\corauthref{cor}},
\address[mit]{Department of Aeronautics and Astronautics,
MIT, 77 Mass Ave, Cambridge, MA 02139, USA}
\corauth[cor]{Corresponding author.}\ead{qiqi@mit.edu}

\begin{abstract}
This paper describes a forward algorithm and an adjoint algorithm for
computing sensitivity derivatives in chaotic dynamical systems, such
as the Lorenz attractor.  The algorithms compute the derivative
of long time averaged ``statistical'' quantities to infinitesimal
perturbations of the system parameters.
The algorithms are demonstrated on the Lorenz attractor.
We show that sensitivity derivatives of statistical quantities can be
accurately estimated using a single, short trajectory (over a time interval
of 20) on the Lorenz attractor.
\end{abstract}

\end{frontmatter}

\section{Introduction}
\label{s:intro}

Computational methods for sensitivity analysis
is a powerful tool in modern computational science
and engineering.  These methods calculate the derivatives of output quantities
with respect to input parameters in computational simulations.
There are two types of algorithms for computing sensitivity
derivatives: the forward algorithms and the adjoint algorithms.  The forward
algorithms are more efficient for computing sensitivity derivatives of many
output quantities to a few input parameters; the adjoint
algorithms are more efficient for computing sensitivity derivatives of a few
output quantities to many input parameters.
Key application of computational methods for sensitivity analysis
include aerodynamic shape optimization \cite{Jameson1988}, adaptive grid
refinement \cite{adaptation}, and data assimilation for weather
forecasting \cite{QJ:QJ49711750206}.

In simulations of chaotic dynamical systems, such as turbulent flows and
the climate system, many output quantities of interest are ``statistical
averages''.  Denote the state of the dynamical system as $x(t)$;
for a function of the state $J(x)$,
the corresponding statistical quantity $\langle J\rangle$ is defined
as an average of $J(x(t))$ over an infinitely long time interval:
\begin{equation} \label{stats}
\langle J\rangle = \lim_{T\rightarrow\infty}
\frac1T \int_0^T J(x(t))\,dt\;,
\end{equation}
For ergodic dynamical systems, a statistical average only depends on the
governing dynamical system, and does not depend on the particular choice of
trajectory $x(t)$.

Many statistical averages, such as the mean aerodynamic forces in
turbulent flow simulations, and the mean global temperature in
climate simulations, are of great scientific and engineering
interest.  Computing sensitivities of these statistical quantities
to input parameters can be useful in many applications.

The differentiability of these statistical
averages to parameters of interest as been established through the recent
developments in the Linear Response Theory for dissipative chaos
\cite{springerlink:10.1007/s002200050134}\cite{0951-7715-22-4-009}.
A class of chaotic dynamical systems,
known as ``quasi-hyperbolic'' systems, has been proven to have statistical
quantities that are differentiable with respect to small
perturbations.  These systems include the Lorenz attractor, and
possibly many systems of engineering interest, such as turbulent
flows.

Despite recent advances both in Linear Response Theory
\cite{0951-7715-22-4-009} and in numerical methods for sensitivity
computation of unsteady systems \cite{wang_siam08}
\cite{lcoadj_jcp}, sensitivity computation of statistical quantities
in chaotic dynamical systems remains difficult.
A major challenge in computing sensitivities in chaotic dynamical
systems is their sensitivity to the initial condition, commonly
known as the ``butterfly effect''.  The linearized equations, used both in
forward and adjoint sensitivity computations, give rise to solutions
that blow up exponentially.
When a statistical quantity is approximated by a finite time average,
the computed sensitivity derivative of the finite time average diverges
to infinity, instead of converging to the sensitivity derivative of the
statistical quantity \cite{leaclimate}.
Existing methods for computing correct sensitivity derivatives of
statistical quantities usually involve averaging over a large number
of ensemble calculations \cite{leaclimate} \cite{eyinkclimate}.
The resulting high computation cost makes these methods not attractive
in many applications.

This paper outlines a computational method for efficiently
estimating the sensitivity derivative of time averaged statistical
quantities, relying on a single trajectory over a small time interval.
The key idea of our method, inversion of the ``shadow'' operator,
is already used as a tool for proving structural stability
of strange attractors \cite{springerlink:10.1007/s002200050134}.
The key strategy of our method, divide and conquer
of the shadow operator,
is inspired by recent advances in numerical computation of the
Lyapunov covariant vectors \cite{PhysRevLett.99.130601}\cite{TELA:TELA234}.

In the rest of this paper,
Section \ref{s:shadow} describes the ``shadow'' operator,
on which our method is based.  Section \ref{s:invert}
derives the sensitivity analysis algorithm by inverting the shadow
operator.  Section \ref{s:time} introduces a fix to the
singularity of the shadow operator.
Section \ref{s:forward} summarizes the forward sensitivity analysis
algorithm.  Section \ref{s:adjoint} derives the corresponding adjoint
version of the sensitivity analysis algorithm.
Section \ref{s:lorenz} demonstrates both the forward and adjoint algorithms
on the Lorenz attractor.  Section \ref{s:conclusion} concludes
this paper.

The paper uses the following mathematical notation:
Vector fields in the state space (e.g. $f(x)$, $\phi_i(x)$) are
column vectors; gradient of scalar fields (e.g. $\frac{\partial
a_i^x}{\partial x}$) are row vectors; gradient of vector fields
(e.g. $\frac{\partial f}{\partial x}$) are matrices with each row being
a dimension of $f$, and each column being a dimension of $x$.  The
($\cdot$) sign is used to identify matrix-vector products
or vector-vector inner products.  For a trajectory $x(t)$ satisfying
$\frac{dx}{dt} = f(x)$ and a
scalar or vector field $a(x)$ in the state space, we often use
$\frac{da}{dt}$ to denote $\frac{da(x(t))}{dt}$.  The chain rule
$\frac{da}{dt} = \frac{da}{dx}\cdot \frac{dx}{dt}
= \frac{da}{dx}\cdot f$ is often used without
explanation.

\section{The ``Shadow Operator''}
\label{s:shadow}

For a smooth, uniformly bounded $n$ dimensional vector field
$\delta x(x)$, defined
on the $n$ dimensional state space of $x$.  The following transform
defines a slightly ``distorted'' coordinates of the state space:
\begin{equation} \label{wrap0}
x'(x) = x + \epsilon\, \delta x(x)
\end{equation}
where $\epsilon$ is a small real number.  Note that for an infinitesimal
$\epsilon$, the following relation holds:
\begin{equation} \label{wrap}
x'(x) - x = \epsilon\, \delta x(x) = \epsilon\, \delta x(x') + O(\epsilon^2)
\end{equation}

\begin{figure}[htb] \centering
\includegraphics[width=3.2in,trim=55 45 45 45,clip]{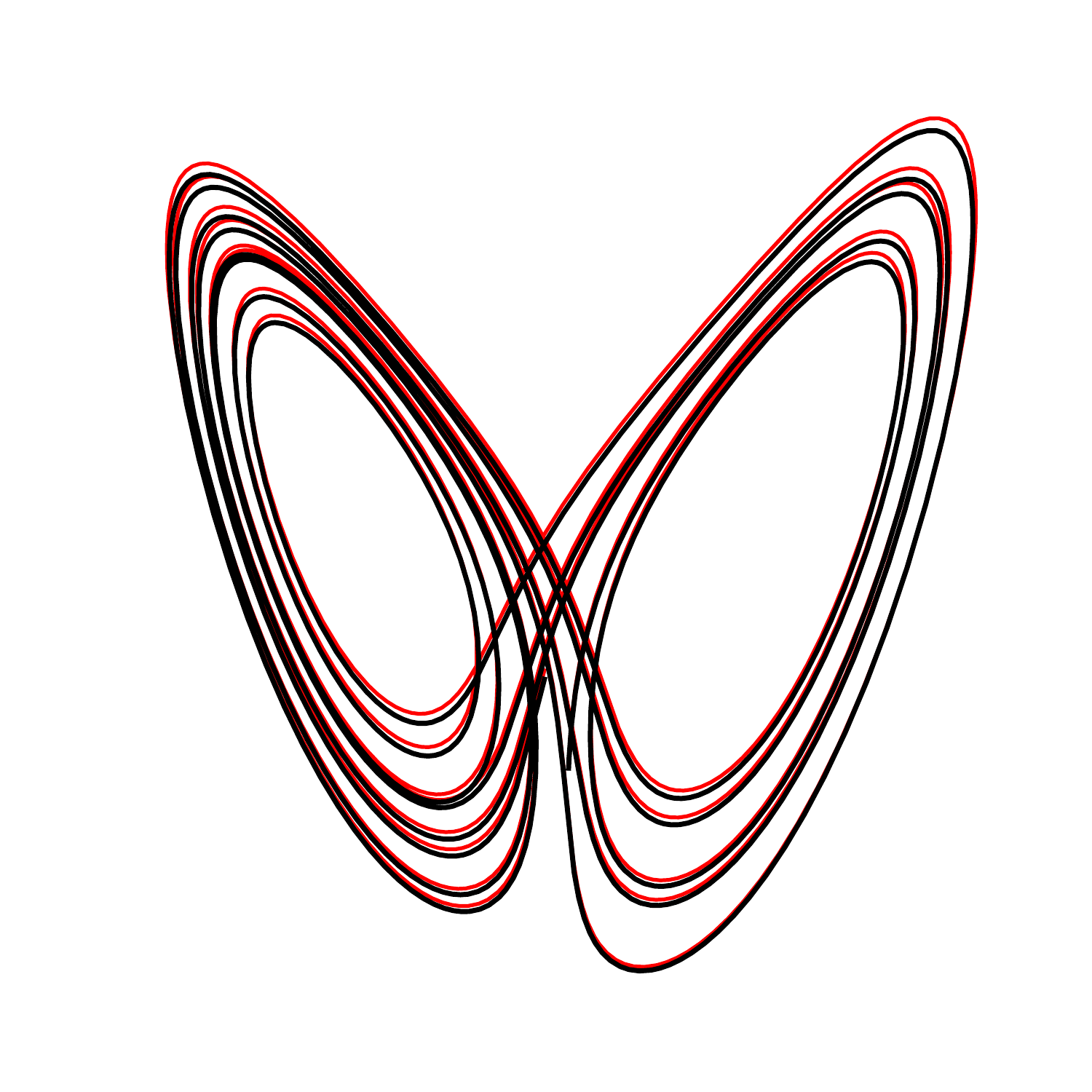}
\caption{A trajectory of the Lorenz attractor under a shadow coordinate
transform.  The black trajectory shows $x(t)$, and the red
trajectory shows $x'(t)$.  The perturbation $\epsilon\,\delta x$
shown corresponds to an infinitesimal change in the parameter $r$, and
is explained in detail in Section \ref{s:lorenz}.}
\label{f:shadow}
\end{figure}

We call the transform from $x$ to $x'$ as a ``shadow coordinate transform''.
In particular, consider a trajectory $x(t)$ and the corresponding
transformed trajectory $x'(t)=x'(x(t))$.  For a small $\epsilon$, the
transformed trajectory $x'(t)$ would ``shadow'' the original
trajectory $x(t)$, i.e., it stays uniformly close to $x(t)$ forever.
Figure \ref{f:shadow} shows an example of a trajectory and its shadow.

Now consider a trajectory $x(t)$ satisfying an ordinary differential
equation
\begin{equation}\label{ode}
\dot{x} = f(x) \;,
\end{equation}
with a smooth vector field $f(x)$ as a function of $x$.
The same trajectory in the transformed ``shadow'' coordinates $x'(t)$
do not satisfy the same differential equation.  Instead,
from Equation (\ref{wrap}), we obtain
\begin{equation}\label{wrapode}
\begin{split}
\dot{x'} &= f(x) + \epsilon\, \frac{\partial \delta x}{\partial x} \cdot f(x) \\
         &= f(x') - \epsilon\, \frac{\partial f}{\partial x} \cdot \delta x(x')
                 + \epsilon\, \frac{\partial \delta x}{\partial x} \cdot f(x') +
                 O(\epsilon^2)
\end{split}
\end{equation}
In other words, the shadow trajectory $x'(t)$ satisfies a
slightly perturbed equation
\begin{equation}\label{dfode}
\dot{x'} = f(x') + \epsilon\, \delta f(x') + O(\epsilon^2)
\end{equation}
where the perturbation $\delta f$ is
\begin{equation}\label{Sf}
\begin{split}
\delta f(x) &= -\frac{\partial f}{\partial x} \cdot \delta x(x)
             + \frac{\partial \delta x}{\partial x} \cdot f(x)  \\
            &= -\frac{\partial f}{\partial x} \cdot \delta x(x)
             + \frac{d \delta x}{dt}\\
           :&= (S_f \delta x)(x)
\end{split}
\end{equation}

For a given differential equation $\dot{x} = f(x)$,
Equation (\ref{Sf}) defines a linear operator $S_f : \delta x
\Rightarrow \delta f$.  We call $S_f$ the ``{\bf Shadow
Operator}'' of $f$.  For any smooth vector field $\delta x(x)$
that defines a slightly distorted ``shadow'' coordinate system in
the state space, $S_f$ determines a
unique smooth vector field $\delta f(x)$ that defines a perturbation
to the differential equation.
Any trajectory of the original differential equation would
satisfy the perturbed equation in the distorted coordinates.

Given an ergodic dynamical system $\dot{x} = f(x)$, and a pair
$(\delta x, \delta f)$ that satisfies $\delta f =
S_f \delta x$, $\delta x$ determines the {\bf sensitivity of
statistical quantities} of the dynamical system to an infinitesimal
perturbation $\epsilon\delta f$.
Let $J(x)$ be a smooth scalar function of the state, consider
the statistical average $\langle J\rangle$ as defined in Equation
(\ref{stats}).  The sensitivity derivative of $\langle J\rangle$
to the infinitesimal perturbation $\epsilon\,\delta f$ is by definition
\begin{equation} \label{presens}
\frac{d \langle J\rangle}{d\epsilon}
= \lim_{\epsilon\rightarrow 0}\frac{1}{\epsilon}
\left(\lim_{T\rightarrow\infty} \frac1T \int_0^T J(x'(t))\,dt
-\lim_{T\rightarrow\infty} \frac1T \int_0^T J(x(t))\,dt \right)
\end{equation}
where by the ergodicity assumption, the statistical average of the perturbed
system can be computed by averaging over $x'(t)$, which satisfies
the perturbed governing differential equation.
Continuing from Equation (\ref{presens}),
\begin{equation} \label{sens}
\begin{split}
\frac{d \langle J\rangle}{d\epsilon}
&= \lim_{\epsilon\rightarrow 0} \lim_{T\rightarrow\infty}
\frac1T \int_0^T \frac{J(x'(t))-J(x(t))}{\epsilon}\:dt \\
&= \lim_{T\rightarrow\infty} \lim_{\epsilon\rightarrow 0}
\frac1T \int_0^T \frac{J(x'(t))-J(x(t))}{\epsilon}\:dt \\
&= \lim_{T\rightarrow\infty}
\frac1T \int_0^T \frac{\partial J}{\partial x} \cdot \delta x\:dt 
= \left\langle
\frac{\partial J}{\partial x} \cdot \delta x \right\rangle \;.
\end{split}
\end{equation}
Equation (\ref{sens}) represents the sensitivity derivative of
a statistical quantity $\langle J\rangle$ to the size of a
perturbation $\epsilon\delta f$.  There are two subtle points here:
\begin{itemize}
\item The two limits $\lim_{\epsilon\rightarrow 0}$
and $\lim_{T\rightarrow\infty}$ can commute with each other
for the following reason: The two trajectories $x'(t)$
and $x(t)$ stay {\bf uniformly}
close to each other {\bf forever}; therefore,
\begin{equation} \frac{J(x'(t))-J(x(t))}{\epsilon}
   \overset{\epsilon\rightarrow 0}\longrightarrow 
   \frac{\partial J}{\partial x} \cdot \delta x
\end{equation}
uniformly for all $t$.  Consequently,
\begin{equation}
   \frac1T \int_0^T \frac{J(x'(t))-J(x(t))}{\epsilon}\:dt
   \overset{\epsilon\rightarrow 0}\longrightarrow 
   \frac1T \int_0^T \frac{\partial J}{\partial x} \cdot \delta x\;dt
\end{equation}
uniformly for all $T$.  Thus the two limits commute.
\item The two trajectories $x'(t)$ and $x(t)$ start at two
specially positioned pair of initial conditions
$x'(0) = x(0) + \epsilon\, \delta x(x(0))$.
Almost any other pair of initial conditions (e.g. $x'(0)=x(0)$)
would make the two trajectories diverge as a result of the
``butterfly effect''.
They would not stay uniformly close to each other, and the limits
$\lim_{\epsilon\rightarrow 0}$ and $\lim_{T\rightarrow\infty}$
would not commute.
\end{itemize}
Equation (\ref{sens}) represents the sensitivity derivative of the
statistical quantity $\langle J\rangle$ to the infinitesimal perturbation
$\epsilon\,\delta f$ as another statistical quantity
$\langle\frac{\partial J}{\partial x} \cdot \delta x\rangle$.  We
can compute it by averaging $\frac{\partial J}{\partial x} \cdot \delta
x$ over a sufficiently long trajectory,
provided that $\delta x = S^{-1} \delta f$ is known along the trajectory.
The next section describes how to numerically compute
$\delta x = S^{-1} \delta f$ for a given $\delta f$.

\section{Inverting the Shadow Operator}
\label{s:invert}

Perturbations to input parameters can often be represented as
perturbations to the dynamics.  Consider a differential equation
$\dot{x} = f(x, s_1, s_2, \ldots, s_m)$ parameterized by $m$ input
variables, an infinitesimal perturbation in a input parameter
$s_j \rightarrow s_j + \epsilon$ can be represented
as a perturbation to the dynamics
$\epsilon\, \delta f = \epsilon\, \frac{df}{ds_j}$.

Equation (\ref{sens}) defines the sensitivity derivative
of the statistical quantity $\langle J\rangle$ to an infinitesimal
perturbation $\epsilon\,\delta f$, provided that a $\delta x$ can be found
satisfying $\delta f = S_f \delta x$, where $S_f$ is the shadow operator.
To compute the sensitivity by evaluating Equation (\ref{sens}),
one must first numerically invert $S_f$ for a given $\delta f$
to find the corresponding $\delta x$.

The key ingredient of numerical inversion of $S_f$ is
the Lyapunov spectrum decomposition.
This decomposition can be efficiently computed
numerically \cite{TELA:TELA234} \cite{PhysRevLett.99.130601}.
In particular, we focus on the case when
the system $\dot{x} = f(x)$ has distinct Lyapunov
exponents.   Denote the Lyapunov covariant vectors as
$ \phi_1(x), \phi_2(x), \ldots,\phi_n(x) $.
Each $\phi_i$ is a vector field in the state space satisfying
\begin{equation} \label{lcv}
\frac{d}{dt}\phi_i(x(t)) = \frac{\partial f}{\partial x}\cdot \phi_i(x(t)) -
\lambda_i \phi_i(x(t))
\end{equation}
where
$\lambda_1,\lambda_2,\ldots,\lambda_n$ are the Lyapunov exponents in
decreasing order.

The Lyapunov spectrum decomposition enables a divide and conquer
strategy for computing $\delta x = S_f^{-1} \delta f$.
For any $\delta f(x)$ and every point $x$ on the attractor,
both $\delta x(x)$ and $\delta f(x)$ can be decomposed into the Lyapunov
covariant vector directions almost everywhere, i.e.
\begin{equation} \label{ax}
\delta x(x) = \sum_{i=1}^n a^x_i(x)\, \phi_i(x)\;,
\end{equation}
\begin{equation} \label{af}
\delta f(x) = \sum_{i=1}^n a^f_i(x)\, \phi_i(x)\;,
\end{equation}
where $a^x_i$ and $a^f_i$ are scalar fields in the state space.
From the form of $S_f$ in Equation (\ref{Sf}), we obtain
\begin{equation} \label{Sftmp1}
\begin{split}
S_f (a^x_i \phi_i)
 =& -\frac{\partial f}{\partial x}\cdot (a^x_i(x) \phi_i(x))
  + \frac{d}{dt}  (a^x_i(x)\, \phi_i(x)) \\
 =& -a^x_i(x)\: \frac{\partial f}{\partial x}\cdot \phi_i(x)
  + \frac{d\, a^x_i(x)}{dt}\, \phi_i(x) + a^x_i(x)\: \frac{d\, \phi_i(x)}{dt}  \;.
\end{split}
\end{equation}
By substituting Equation (\ref{lcv}) into the last term of
Equation (\ref{Sftmp1}), we obtain
\begin{equation} \label{Sfi1}
S_f (a^x_i \phi_i)
 = \left(\frac{d a^x_i(x)}{dt} - \lambda_i \,
 a^x_i(x)\right)\,\phi_i(x)\;,
\end{equation}
By combining Equation (\ref{Sfi1}) with Equations
(\ref{ax}), (\ref{af}) and the linear relation
$\delta f = S_f \delta x$, we finally obtain
\begin{equation} \label{Sfi}
\delta f = \sum_{i=1}^n S_f (a^x_i \phi_i)
 = \sum_{i=1}^n\; \underbrace{\left(\frac{d a^x_i}{dt} - \lambda_i \,
 a^x_i\right)}_{\displaystyle a^f_i}\,\phi_i\;,
\end{equation}

Equations (\ref{Sfi1}) and (\ref{Sfi}) are useful for two reasons:
\begin{enumerate}
\item They indicate that the Shadow Operator $S_f$, applied to a scalar field
$a^x_i(x)$ multiple of $\phi_i(x)$, generates another scalar field
$a^f_i(x)$ multiple of the same vector field $\phi_i(x)$.
Therefore, one can compute
$S_f^{-1} \delta f$ by first decomposing $\delta f$ as in Equation
(\ref{af}) to obtain the $a^f_i$.  If each $a_i^x$ can be calculated
from the corresponding $a_i^f$, then $\delta x$ can be computed
with Equation (\ref{ax}), completing the inversion.
\item
It defines a scalar ordinary
differential equation that governs the relation between $a^x_i$ and
$a^f_i$ along a trajectory $x(t)$:
\begin{equation} \label{Sfode0}
\frac{d a^x_i(x)}{dt} = a^f_i(x) + \lambda_i \, a^x_i(x)
\end{equation}
This equation can be used to obtain $a^x_i$ from $a^f_i$ along a
trajectory, thereby filling the gap in the inversion procedure of $S_f$
outlined above.
For each positive Lyapunov exponent $\lambda_i$, one can integrate the
ordinary differential equation
\begin{equation} \label{Sfode}
\frac{d \check{a}^x_i}{dt} = \check{a}^f_i + \lambda_i \, \check{a}^x_i
\end{equation}
backwards in time from an arbitrary terminal condition, and the
difference between $\check{a}^x_i(t)$ and the desired $a^x_i(x)$
will decrease exponentially.
For each negative Lyapunov exponent $\lambda_i$, Equation
(\ref{Sfode}) can be integrated forward in time from an arbitrary initial
condition, and $\check{a}^x_i(t)$ will converge exponentially
to the desired $a^x_i(x)$.
For a zero Lyapunov exponent $\lambda_i=0$, Section \ref{s:time}
introduces a solution.
\end{enumerate}

\section{Time Dilation and Compression}
\label{s:time}

There is a fundamental problem in the inversion method derived in
Section \ref{s:invert}:  $S_f$ is not invertible for certain $\delta f$.
This can be shown with the following analysis:
Any continuous time
dynamical system with a non-trivial attractor must have a zero Lyapunov
exponent $\lambda_{n_0}=0$.  The corresponding Lyapunov covariant
vector is $\phi_{n_0}(x) = f(x)$.  This can be verified by substituting
$\lambda_i=0$ and $\phi_i=f$ into Equation (\ref{lcv}).
For this $i=n_0$, Equations (\ref{Sfode}) becomes
\begin{equation} \label{time1}
a^f_{n_0}(x) = \frac{d a^x_{n_0}(x)}{dt}
\end{equation}
By taking an infinitely long time average on both sides of Equation
(\ref{time1}), we obtain
\begin{equation} \label{zeroLyapRange}
\left\langle a^f_{n_0}(x) \right\rangle
= \lim_{T\rightarrow\infty} \frac{a^x_{n_0}(x(T)) - a^x_{n_0}(x(0))}{T} = 0\;,
\end{equation}

Equation (\ref{zeroLyapRange}) implies that for any $\delta f = S_f
\delta x$, the $i=n_0$ component of its Lyapunov decomposition
(as in Equation (\ref{af})) must satisfy $\langle a^f_{n_0}(x) \rangle = 0$.
Any $\delta f$ that do not satisfy this linear relation, e.g.
$\delta f \equiv f$, would not be in the range space of $S_f$.  Thus the
corresponding $\delta x = S_f^{-1}\delta f$ does not exist.

Our solution to the problem is complementing $S_f$ with
a ``global time dilation and compression'' constant $\eta$, whose effect
produces a $\delta f$ that is outside the range space of $S_f$.
We call $\eta$ a time dilation constant for short.
The combined effect of a time dilation constant and a shadow
transform could produce all smooth perturbations $\delta f$.

The idea comes from the fact that for a constant $\eta$, the time dilated or
compressed system $\dot{x} = (1 + \epsilon\, \eta) f(x)$
has exactly the same statistics
$\langle J\rangle$, as defined in Equation (\ref{stats}),
as the original system $\dot{x} = f(x)$.
Therefore, the perturbation in any $\langle
J\rangle$ due to any $\epsilon\, \delta f$ is equal to the
perturbation in $\langle J\rangle$ due to
$\epsilon\, (\eta f(x) + \delta f(x))$.
Therefore, the sensitivity derivative to $\delta f$
can be computed if we can find a
$\delta x$ that satisfies $S_f \delta x = \eta f(x) + \delta f(x)$
for some $\eta$.

We use the ``free'' constant $\eta$ to put $\eta f(x) + \delta f(x)$ into
the range space of $S_f$.
By substituting $\eta f(x) + \delta f(x)$ into the constraint Equation
(\ref{zeroLyapRange}) that identifies the range space of $S_f$,
the appropriate $\eta$ must satisfy the following equation
\begin{equation} \label{eta}
   \eta + \langle a^f_{n_0} \rangle = 0\;,
\end{equation}
which we use to numerically compute $\eta$.

Once the appropriate time dilation constant $\eta$ is computed,
$\eta f(x) + \delta f(x)$ is in the range space of $S_f$.
We use the procedure in Section
\ref{s:invert} to compute $\delta x = S_f^{-1} (\eta f + \delta f)$,
then use Equation (\ref{sens}) to compute the desired
sensitivity derivative $d\langle J\rangle/d\epsilon$.
The addition of $\eta f$ to $\delta f$ affects Equation (\ref{Sfode})
only for $i=n_0$, making it
\begin{equation} \label{Sfoden0}
\frac{d a^x_{n_0}(x)}{dt} = a^f_{n_0}(x) + \eta\;.
\end{equation}
Equation (\ref{Sfoden0}) indicates that $a^x_{n_0}$ can be computed
by integrating the right hand side along the trajectory.

The solution to Equation (\ref{Sfoden0}) admits an arbitrary additive
constant.  The effect of this arbitrary constant is the following:
By substituting Equations (\ref{ax}) into Equation
(\ref{sens}), the contribution from the $i = n_0$ term of $\delta x$
to $d\langle J\rangle/d\epsilon$ is
\begin{equation} \label{zeroLyapSens}
   \lim_{T\rightarrow\infty}\frac1T \int_0^T a^f_{n_0} \frac{dJ}{dt}\,dt
\end{equation}
Therefore, any constant addition to $a^f_{n_0}$ vanishes as
$T\rightarrow\infty$.  Computationally,
however, Equation (\ref{sens}) must be approximated by a finite time average.
We find it beneficial to adjust the level of $a^f_{n_0}$ to
approximately $\langle a^f_{n_0} \rangle = 0$, in order
to control the error due to finite time averaging.

\section{The Forward Sensitivity Analysis Algorithms}
\label{s:forward}

For a given $\dot{x} = f(x)$, $\delta f$ and $J(x)$,
the mathematical developments in Sections \ref{s:invert} and \ref{s:time}
are summarized into Algorithm \ref{alg:1} for
computing the sensitivity derivative
$d\delta \langle J\rangle/d\epsilon$ as in Equation
(\ref{sens}).

\begin{algorithm}
\caption{The Forward Sensitivity Analysis Algorithm}
\label{alg:1}
\begin{enumerate}
\item \label{alg1:step1}
Choose a ``spin-up buffer time'' $T_B$, and an
``statistical averaging time'' $T_A$.
$T_B$ should be much longer than $1/|\lambda_i|$ for all nonzero
Lyapunov exponent $\lambda_i$, so that the solutions of Equation
(\ref{Sfode}) can reach $a_i^x$ over a time span of $T_B$.
$T_A$ should be much longer than the decorrelation time
of the dynamics, so that one can accurately approximate a statistical
quantity by averaging over $[0, T_A]$.
\item \label{alg1:step2}
Obtain an initial condition on the attractor at $t=-T_B$, e.g.,
by solving $\dot{x} = f(x)$ for a sufficiently long time span,
starting from an arbitrary initial condition.
\item \label{alg1:step3}
Solve $\dot{x} = f(x)$ to obtain a trajectory
$x(t), t\in[-T_B,T_A+T_B]$; compute the Lyapunov exponents
$\lambda_i$ and the Lyapunov covariant vectors
$\phi_i(x(t))$ along the trajectory, e.g., using algorithms in
\cite{TELA:TELA234} and \cite{PhysRevLett.99.130601}.
\item \label{alg1:step4}
Perform the Lyapunov spectrum decomposition of
$\delta f(x)$ along the trajectory $x(t)$ to
obtain $a^f_i(x), i=1,\ldots,n$ as in Equation (\ref{af}).
\item \label{alg1:step5}
Compute the global time dilation constant $\eta$
using Equation (\ref{eta}).
\item \label{alg1:step6}
Solve the differential equations (\ref{Sfode}) to obtain
$a^x_i$ over the time interval $[0, T_A]$.  The equations with
positive $\lambda_i$ are solved backward in time from $t=T_A+T_B$
to $t=0$; the ones with negative $\lambda_i$ are solved forward in
time from $t=-T_B$ to $t=T_A$.  For $\lambda_{n_0} = 0$,
Equation (\ref{Sfoden0}) is integrated, and the
mean of $a^x_{n_0}$ is set to zero.
\item \label{alg1:step7}
Compute $\delta x$ along the trajectory $x(t), t\in[0,T_A]$
with Equation (\ref{ax}).
\item \label{alg1:step8}
Compute $d\langle J\rangle/d\epsilon$ using Equation (\ref{stats})
by averaging over the time interval $[0,T_A]$.
\end{enumerate}
\end{algorithm}

The preparation phase of the algorithm (Steps
\ref{alg1:step1}-\ref{alg1:step3})
computes a trajectory and the Lyapunov spectrum decomposition
along the trajectory.  The algorithm then starts by decomposing $\delta f$
(Step \ref{alg1:step4}), followed by computing $\delta x$
(Steps \ref{alg1:step5}-\ref{alg1:step7}), and
finally computing $d\langle J\rangle/d\epsilon$ (Step \ref{alg1:step8}).
The sensitivity derivative of
many different statistical quantities $\langle J_1\rangle, \langle J_2\rangle,
\ldots$ to a single $\delta f$ can be computed by only repeating the
last step of the algorithm.
Therefore, this is a ``forward'' algorithm in the sense that
it efficiently computes sensitivity of multiple output quantities to
a single input parameter (the size of perturbation
$\epsilon\,\delta f$).  We will see that this is in sharp
contrast to the ``adjoint'' algorithm described in Section
\ref{s:adjoint}, which efficiently computes the
sensitivity derivative of one output statistical quantity
$\langle J\rangle$ to many
perturbations $\delta f_1, \delta f_2, \ldots$.

It is worth noting that the $\delta x$ computed using Algorithm \ref{alg:1}
satisfies the forward tangent equation
\begin{equation} \label{alltan}
    \dot{\delta x} = \frac{\partial f}{\partial x}\cdot \delta x
                   + \eta\, f + \delta f
\end{equation}
This can be verified by taking derivative of Equation (\ref{ax}),
substituting Equations (\ref{Sfode}) and (\ref{Sfoden0}), then using
Equation (\ref{af}).  However, $\delta x$ must satisfy both an initial
condition and a terminal condition, making it difficult to solve
with conventional time integration methods.
In fact, Algorithm \ref{alg:1} is equivalent
to splitting $\delta x$ into stable, neutral and unstable
components, corresponding to positive, zero and negative Lyapunov
exponents; then solving Equation (\ref{alltan}) separately for each
component in different time directions.
This alternative version of the forward sensitivity computation
algorithm could be
useful for large systems to avoid computation of all the Lyapunov
covariant vectors.

\section{The Adjoint Sensitivity Analysis Algorithm}
\label{s:adjoint}

The adjoint algorithm starts by trying to find an {\bf adjoint vector field}
$\hat{f}(x)$, such that the sensitivity derivative of
the given statistical quantity $\langle J\rangle$ to any infinitesimal
perturbation $\epsilon\, \delta f$ can be represented as
\begin{equation} \label{fadj}
   \frac{d\langle J\rangle}{\epsilon}
 = \left\langle \hat{f\,}^T\cdot \delta f \right\rangle
\end{equation}
Both $\hat{f}$ in Equation (\ref{fadj}) and
$\frac{\partial J}{\partial x}$ in Equation
(\ref{sens}) can be decomposed into linear combinations of the \emph{adjoint
Lyapunov covariant vectors}
almost everywhere on the attractor:
\begin{equation} \label{bf}
\hat{f}(x) = \sum_{i=1}^n {\hat a}^f_i(x)\, \psi_i(x)\;,
\end{equation}
\begin{equation} \label{bx}
\frac{\partial J}{\partial x}^T = \sum_{i=1}^n {\hat a}^x_i(x)\, \psi_i(x)\;,
\end{equation}
where
the adjoint Lyapunov covariant vectors $\psi_i$ satisfy
\begin{equation} \label{alcv}
-\frac{d}{dt}\psi_i(x(t)) = \frac{\partial f}{\partial x}^T\cdot \psi_i(x(t)) -
\lambda_i \psi_i(x(t))
\end{equation}
With proper normalization,
the (primal) Lyapunov covariant vectors $\phi_i$ and the adjoint Lyapunov
covariant vectors $\psi_i$ have the following conjugate relation:
\begin{equation} \label{conjugate}
\psi_i(x)^T \cdot \phi_j(x) \equiv
\begin{cases} 0\;, & i\ne j \\ 1\;, & i = j \end{cases}
\end{equation}
i.e., the $n\times n$ matrix formed by all the $\phi_i$ and the
$n\times n$ matrix formed by all the $\psi_i$ are
the transposed inverse of each
other at every point $x$ in the state space.

By substituting Equations (\ref{ax}) and (\ref{bx}) into Equation
(\ref{sens}), and using the conjugate relation in Equation
(\ref{conjugate}), we obtain
\begin{equation} \label{fadj2}
   \frac{d\langle J\rangle}{d\epsilon}
 = \sum_{i=1}^n \left\langle {\hat a}_i^x a_i^x \right\rangle
\end{equation}
Similarly, by combining Equations (\ref{fadj}), (\ref{af}), (\ref{bf}) and
(\ref{conjugate}), it can be shown that $\hat{f}$ satisfies Equation
(\ref{fadj}) if and only if
\begin{equation} \label{fadj3}
   \frac{d\langle J\rangle}{d\epsilon}
 = \sum_{i=1}^n \left\langle {\hat a}_i^f a_i^f \right\rangle
\end{equation}
Comparing Equations (\ref{fadj2}) and (\ref{fadj3}) leads to the
following conclusion: Equation (\ref{fadj}) can be satisfied by finding
${\hat a}_i^f$ that satisfy
\begin{equation} \label{fadja}
  \left\langle {\hat a}_i^f a_i^f \right\rangle
= \left\langle {\hat a}_i^x a_i^x \right\rangle\;,\quad i=1,\ldots,n
\end{equation}

The ${\hat a}_i^f$ that satisfies Equation (\ref{fadja}) can
be found using the relation between $a_i^f$ and $a_i^x$ in Equation
(\ref{Sfode0}).  By multiplying $\hat{a}_i^f$ on both sides of Equation
(\ref{Sfode0}) and integrate by parts in time, we obtain
\begin{equation} \label{afaxadjtmp1}
\frac1T \int_0^T {\hat a}_i^f a_i^f dt
= \left.\frac{{\hat a}_i^f a_i^x}T \right|_0^T
- \frac1T \int_0^T \left(\frac{d {\hat a}_i^f}{dt}
              + \lambda_i \, {\hat a}_i^f\right) a^x_i\;dt
\end{equation}
for $i\ne n_0$.
Through apply the same technique to Equation (\ref{Sfoden0}), we obtain
for $i=n_0$
\begin{equation} \label{afaxadjtmp2}
\frac1T \int_0^T {\hat a}_{n_0}^f a_{n_0}^f dt
= \left.\frac{{\hat a}_{n_0}^f a_{n_0}^x}T \right|_0^T
- \frac1T \int_0^T \frac{d {\hat a}_{n_0}^f}{dt}\, a^x_{n_0} dt
+ \frac1T \int_0^T \eta\, \hat{a}^f_{n_0} dt
\end{equation}

If we set $\hat{a}_i^f$ to satisfy the following relations
\begin{equation} \label{afaxadj0}
\begin{aligned}
-\frac{d {\hat a}_i^f(x)}{dt} &= \hat{a}_i^x(x) +
 \lambda_i\, {\hat a}_i^f(x)\;,
 & i\ne n_0 \;,\\
-\frac{d {\hat a}_i^f(x)}{dt} &= \hat{a}_i^x(x) \;,\quad 
 \langle\hat{a}_i^f\rangle = 0\;,\quad
 & i = n_0 \;,
\end{aligned}
\end{equation}
then Equations (\ref{afaxadjtmp1}) and (\ref{afaxadjtmp2}) become
\begin{equation} \label{fadjaver1}
\begin{aligned}
\frac1T \int_0^T {\hat a}_i^f a_i^f dt
&= \left.\frac{{\hat a}_i^f a_i^x}T \right|_0^T
+ \frac1T \int_0^T \hat{a}_i^x a_i^x\;dt \;,\quad & i\ne n_0 \\
\frac1T \int_0^T {\hat a}_i^f a_i^f dt
&= \left.\frac{{\hat a}_i^f a_i^x}T \right|_0^T
+ \frac1T \int_0^T \hat{a}_i^x a_i^x\;dt
+ \eta \left(\frac1T \int_0^T \hat{a}^f_{n_0}\,dt
 -\langle\hat{a}^f_{n_0}\rangle\right), & i= n_0 \\
\end{aligned}
\end{equation}
As $T\rightarrow\infty$, both equations reduces to Equation (\ref{fadja}).

In summary, if the scalar fields $\hat{a}_i^f$ satisfy Equation
(\ref{afaxadj0}),
then they also satisfy Equation (\ref{fadjaver1}) and thus
Equation (\ref{fadja}); as a result,
the $\hat{f}$ formed by these $\hat{a}^f$ through Equation (\ref{bf})
satisfies Equation (\ref{fadj}), thus is the desired adjoint vector field.

For each $i\ne n_0$, the scalar field $\hat{a}_i^f$ satisfying
Equation (\ref{afaxadj0}) can be
computed by solving an ordinary differential equations
\begin{equation} \label{afaxadj}
-\frac{d \check{\hat a}_i^f}{dt} = \check{\hat a}_i^x +
 \lambda_i\, \check{\hat a}_i^f\;.
\end{equation}
Contrary to computation of $a_i^x$ through solving Equation (\ref{Sfode}),
the time integration should be forward in time for positive $\lambda_i$, and
backward in time for negative $\lambda_i$, in order for the difference
between $\check{\hat a}_i^f(t)$ and ${\hat a}_i^f(x(t))$ to diminish
exponentially.

The $i=n_0$ equation in Equation (\ref{afaxadj0}) can be directly integrated
to obtain ${\hat a}_{n_0}^f(x)$.
The equation is well defined because the right hand side is mean zero:
\begin{equation}
   \frac1T \int_0^T \hat{a}_{n_0}^f(x(t))\,dt
 = \frac1T \int_0^T \frac{\partial J}{\partial x}\cdot \phi_{n_0}\,dt
 = \frac1T \int_0^T \frac{dJ}{dt}\,dt
 \overset{T\rightarrow\infty}{\longrightarrow} 0\;.
\end{equation}
Therefore, the integral of ${\hat a}_{n_0}^x(x)$ over time, subtracted
by its mean, is the solution ${\hat a}_{n_0}^f(x)$ to
the $i=n_0$ case of Equation (\ref{afaxadj0}).

\begin{algorithm}
\caption{The Adjoint Sensitivity Analysis Algorithm}
\label{alg:2}
\begin{enumerate}
\item \label{alg2:step1}
Choose a ``spin-up buffer time'' $T_B$, and an
``statistical averaging time'' $T_A$.
$T_B$ should be much longer than $1/|\lambda_i|$ for all nonzero
Lyapunov exponent $\lambda_i$, so that the solutions of Equation
(\ref{Sfode}) can reach $a_i^x$ over a time span of $T_B$.
$T_A$ should be much longer than the decorrelation time
of the dynamics, so that one can accurately approximate a statistical
quantity by averaging over $[0, T_A]$.
\item \label{alg2:step2}
Obtain an initial condition on the attractor at $t=-T_B$, e.g.,
by solving $\dot{x} = f(x)$ for a sufficiently long time span,
starting from an arbitrary initial condition.
\item \label{alg2:step3}
Solve $\dot{x} = f(x)$ to obtain a trajectory
$x(t), t\in[-T_B,T_A+T_B]$; compute the Lyapunov exponents
$\lambda_i$ and the Lyapunov covariant vectors
$\phi_i(x(t))$ along the trajectory, e.g., using algorithms in
\cite{TELA:TELA234} and \cite{PhysRevLett.99.130601}.
\item \label{alg2:step4}
Perform the Lyapunov spectrum decomposition of
$(\partial J/\partial x)^T$ along the trajectory $x(t)$ to
obtain $\hat{a}^x_i(x(t)), i=1,\ldots,n$ as in Equation (\ref{bx}).
\item \label{alg2:step5}
Solve the differential equations (\ref{afaxadj}) to obtain
$\hat{a}_i^f(x(t))$ over the time interval $[0, T_A]$.  The equations with
negative $\lambda_i$ are solved backward in time from $t=T_A+T_B$
to $t=0$; the ones with positive $\lambda_i$ are solved forward in
time from $t=-T_B$ to $t=T_A$.  For $i=n_0$, the scalar
$-a^x_{n_0}$ is integrated along the trajectory;
the mean of the integral is subtracted from the integral itself
to obtain $\hat{a}^f_{n_0}$.
\item \label{alg2:step6}
Compute $\hat{f}$ along the trajectory $x(t), t\in[0,T_A]$
with Equation (\ref{bf}).
\item \label{alg2:step7}
Compute $d\langle J\rangle/d\epsilon$ using Equation (\ref{fadj})
by averaging over the time interval $[0,T_A]$.
\end{enumerate}
\end{algorithm}
The above analysis summarizes to Algorithm \ref{alg:2} for
computing the sensitivity derivative derivative of the statistical average
$\langle J\rangle$ to an infinitesimal perturbations $\epsilon\,\delta f$.
The preparation phase of the algorithm (Steps
\ref{alg2:step1}-\ref{alg2:step3}) is exactly the same as in
Algorithm \ref{alg:1}.  These steps compute a trajectory $x(t)$ and the
Lyapunov spectrum decomposition along the trajectory.
The adjoint algorithm then starts by decomposing the derivative vector
$(\partial J/\partial x)^T$ (Step \ref{alg2:step4}),
followed by computing the adjoint vector $\delta f$
(Steps \ref{alg2:step5}-\ref{alg2:step6}), and
finally computing $d\langle J\rangle/d\epsilon$ for a particular $\delta f$.
Note that the sensitivity of the same $\langle J\rangle$ to
many different perturbations $\delta f_1, \delta f_2, \ldots$
can be computed by repeating only the last step of the
algorithm.  Therefore, this is an ``adjoint'' algorithm, in the sense
that it efficiently computes the sensitivity derivatives
of a single output quantity to many input perturbation.

It is worth noting that $\hat{f}$ computed using Algorithm \ref{alg:2}
satisfies the adjoint equation
\begin{equation} \label{alladj}
    -\dot{\hat f} = \frac{\partial f}{\partial x}^T\cdot \hat{f}
                  - \frac{\partial J}{\partial x}
\end{equation}
This can be verified by taking derivative of Equation (\ref{bf}),
substituting Equation (\ref{afaxadj0}), then using
Equation (\ref{bx}).  However, $\hat f$ must satisfy both an initial
condition and a terminal condition, making it difficult to solve
with conventional time integration methods.
In fact, Algorithm \ref{alg:2} is equivalent
to splitting $\hat f$ into stable, neutral and unstable
components, corresponding to positive, zero and negative Lyapunov
exponents; then solving Equation (\ref{alladj}) separately for each
component in different time directions.
This alternative version of the adjoint sensitivity computation
algorithm could be
useful for large systems, to avoid computation of all the Lyapunov
covariant vectors.

\section{An Example: the Lorenz Attractor}

\label{s:lorenz}

We consider the Lorenz attractor $\dot{x}=f(x)$, where
$x = (x_1, x_2, x_3)^T$, and
\begin{equation} \label{lorenz}
f(x) = \left(\begin{array}{c}
   \sigma(x_2-x_1)\\
   x_1(r-x_3) - x_2\\
   x_1 x_2 - \beta x_3
\end{array}\right)
\end{equation}
The ``classic'' parameter values $\sigma=10$, $r=28$, $\beta=8/3$ are used.
Both the forward sensitivity analysis algorithm (Algorithm \ref{alg:1})
and the adjoint sensitivity analysis algorithm (Algorithm \ref{alg:2})
are performed on this system.

We want to demonstrate the computational
efficiency of our algorithm;
therefore, we choose a relatively short statistical averaging interval of
$T_A = 10$, and a spin up buffer period of $T_B=5$.
Only a single trajectory of length $T_A + 2T_B$
on the attractor is required in our algorithm.
Considering that the oscillation period of the Lorenz attractor is around
$1$, the combined trajectory length of $20$ is a reasonable time
integration length for most simulations of chaotic dynamical systems.
In our example, we start the time integration
from $t=-10$ at $x = (-8.67139571762,4.98065219709,25)$, and integrate
the equation to $t=-5$, to ensure that the entire trajectory from
$-T_B$ to $T_A+T_B$ is roughly on the
attractor.  The rest of the discussion in this section are focused on the
trajectory $x(t)$ for $t\in[-T_B,T_A+T_B]$.

\subsection{Lyapunov covariant vectors}

\label{s:decomp}

\begin{figure}[htb]\centering
  \subfloat[The state vector $x$]{\label{f:state}
      \includegraphics[width=0.48\textwidth]{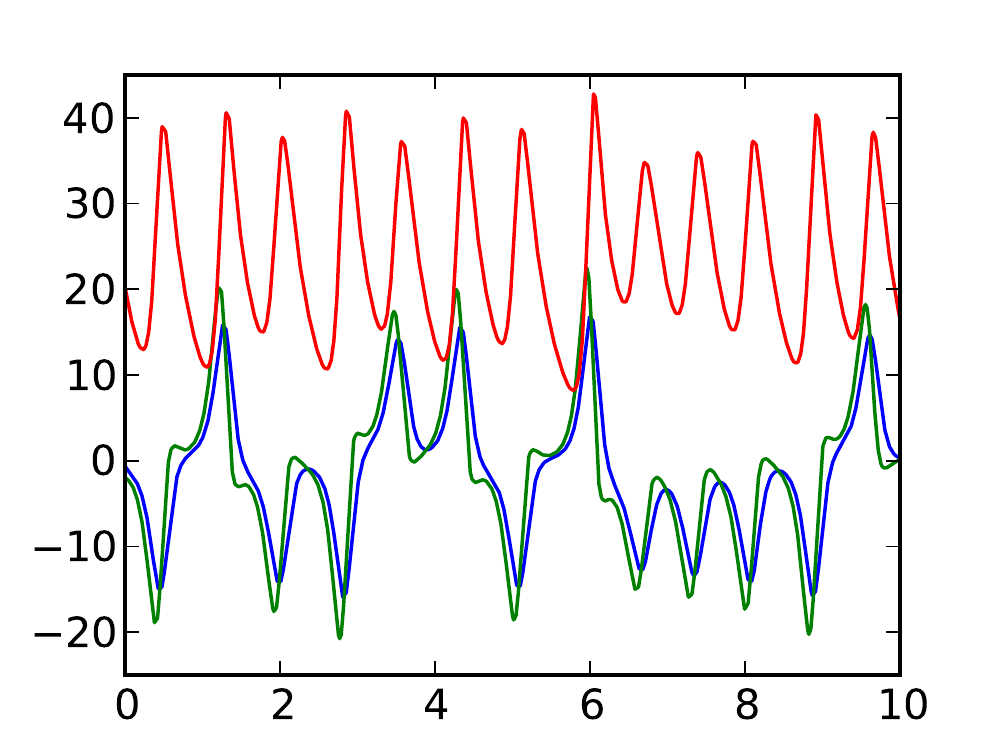}}
  \hspace{0.01\textwidth}
  \subfloat[First Lyapunov covariant vector $\phi_1$]{\label{f:phi1}
      \includegraphics[width=0.48\textwidth]{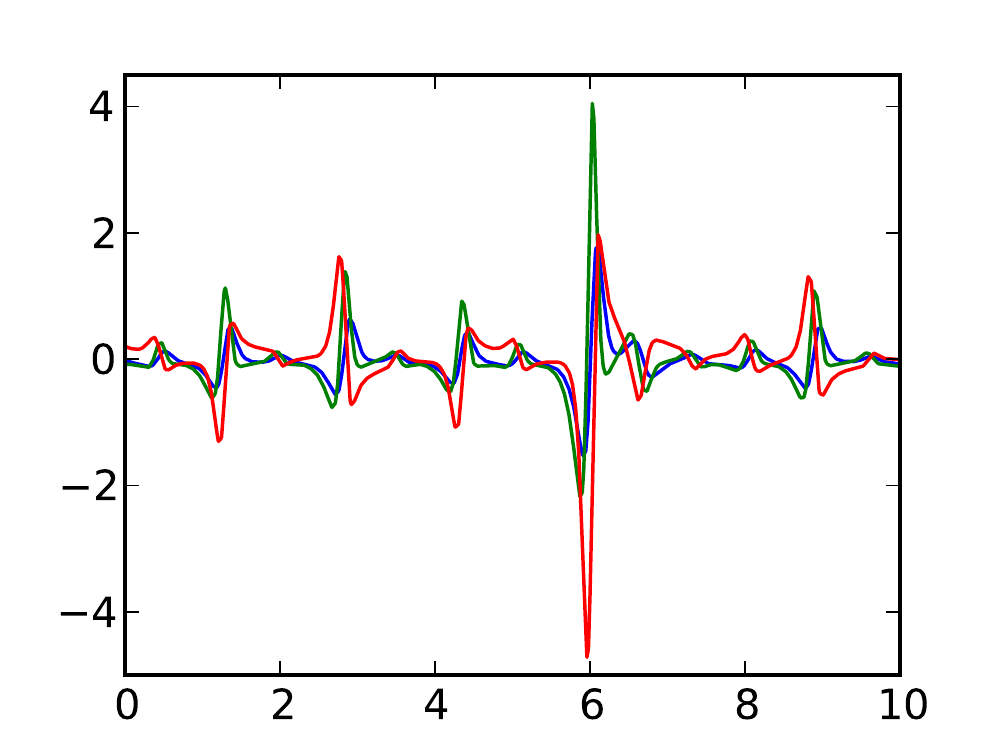}}
  \hspace{0.01\textwidth}
  \subfloat[Second Lyapunov covariant vector $\phi_2$]{\label{f:phi2}
      \includegraphics[width=0.48\textwidth]{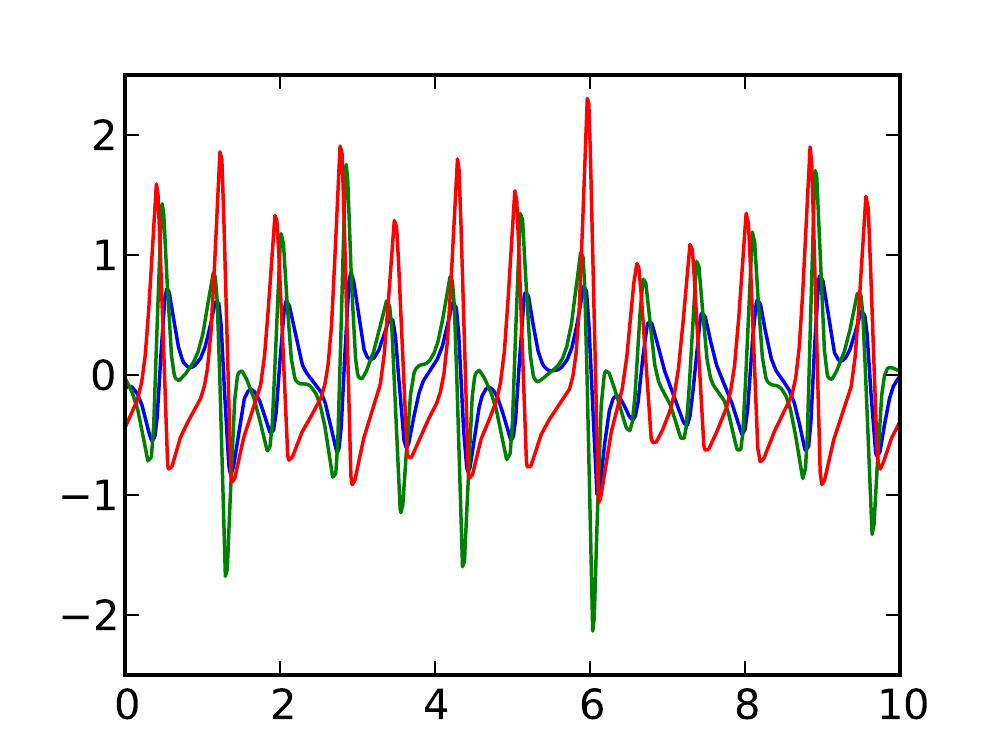}}
  \hspace{0.01\textwidth}
  \subfloat[Third Lyapunov covariant vector $\phi_3$]{\label{f:phi3}
      \includegraphics[width=0.48\textwidth]{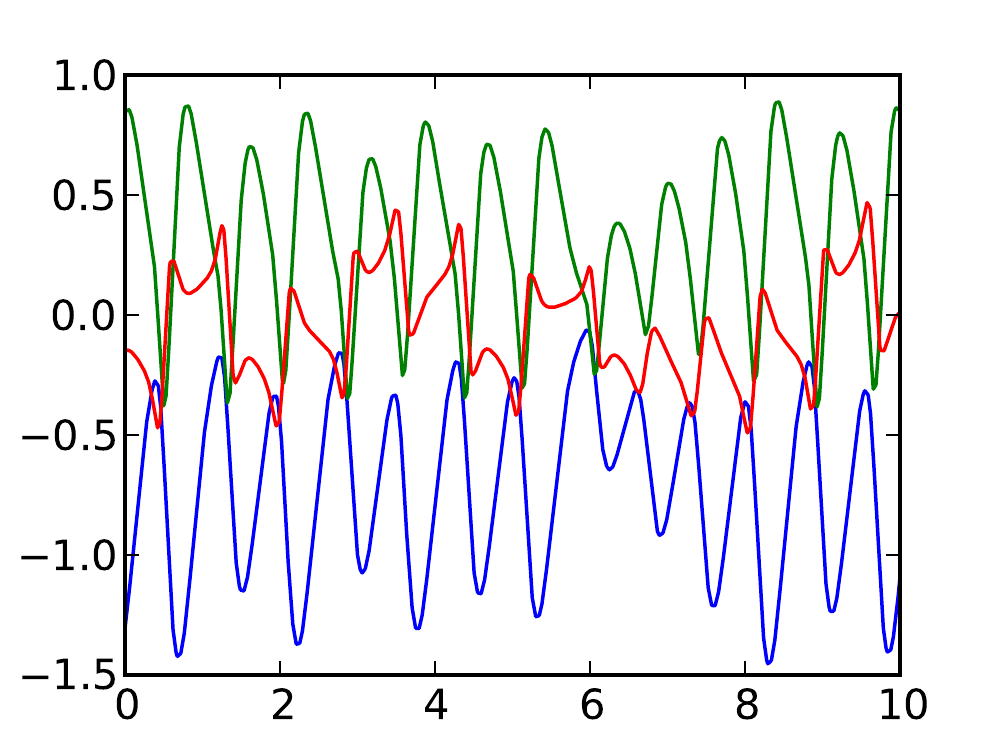}}
  \caption{The Lyapunov covariant vectors of the Lorenz attractor along
  the trajectory $x(t)$ for $t\in [0,10]$.  The x-axes are $t$; the
  blue, green and red lines correspond to the $x_1, x_2$ and $x_3$
  coordinates in the state space, respectively.}
\label{f:decomp}
\end{figure}

The Lyapunov covariant vectors are computed in Step \ref{alg1:step3} of both
Algorithm \ref{alg:1} and Algorithm \ref{alg:2}, over the time
interval $[-T_B,T_A+T_B]$.  These vectors, along with the trajectory
$x(t)$, are shown in Figure \ref{f:decomp}.

The three dimensional
Lorenz attractor has three pairs of Lyapunov exponents and Lyapunov
covariant vectors.  $\lambda_1$ is the only positive
Lyapunov exponent, and $\phi_1$ is computed by integrating the tangent
linear equation
\begin{equation} \label{tangent}
\dot{\tilde{x}} = \frac{\partial f}{\partial x}\cdot \tilde{x}
\end{equation}
forward in time from an arbitrary initial condition at $t=-T_B$.
The first Lyapunov exponent is estimated to be $\lambda_1\approx
0.95$ through a linear regression of $\tilde{x}$ in the log
space.  The first Lyapunov vector is then obtained as
$ \phi_1 = \tilde{x}\, e^{-\lambda_1 t} $.

$\lambda_2=0$ is the vanishing
Lyapunov exponent; therefore, $\phi_2 = \theta\,f(x)$,
where $\theta  = 1 / \sqrt{\langle \|f\|_2^2\rangle}$ is a
normalizing constant that make the mean magnitude of $\phi_2$ equal to 1.

The third Lyapunov exponent $\lambda_3$ is negative.
So $\phi_3$ is computed by integrating the
tangent linear equation (\ref{tangent}) backwards in time from an
arbitrary initial condition at $t=T_A+T_B$.
The third Lyapunov exponent is estimated to be $\lambda_3\approx
-14.6$ through a linear regression of the backward solution
$\tilde{x}$ in the log space.  The third Lyapunov vector is then
obtained as $ \phi_3 = \tilde{x}\, e^{-\lambda_3 t} $.

\subsection{Forward Sensitivity Analysis}

\label{s:lorenzforward}

We demonstrate our forward sensitivity analysis algorithm
by computing the sensitivity
derivative of three statistical quantities $\langle x_1^2\rangle$,
$\langle x_2^2\rangle$ and, $\langle x_3\rangle$ to a small
perturbation in the system parameter $r$ in the Lorenz attractor Equation
(\ref{lorenz}).  The infinitesimal perturbation $r\rightarrow r+\epsilon$
is equivalent to the perturbation
\begin{equation} \label{lorenzdf}
\epsilon \,\delta f = \epsilon\,\frac{\partial f}{\partial r}
     = \epsilon\,( 0, x_1, 0 )^T\;.
\end{equation}

\begin{figure}[htb]\centering
  \subfloat[$\delta f = \dfrac{df}{dr}$]{\label{f:df}
      \includegraphics[width=0.48\textwidth]{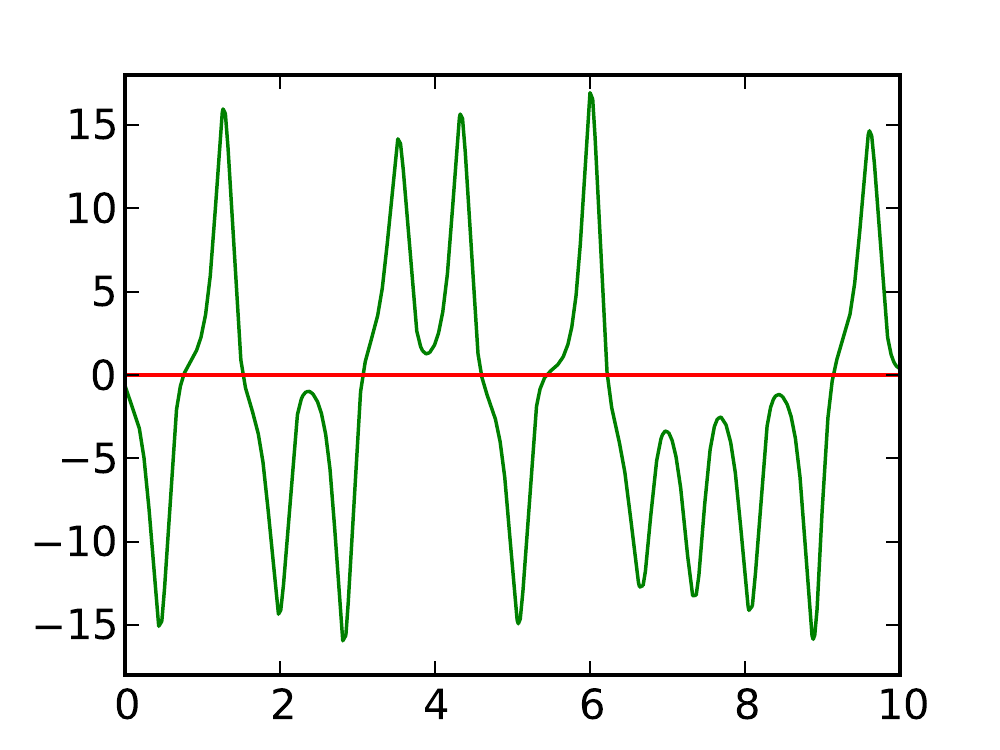}}
  \subfloat[$a^f_i,i=1,2,3$ for the $\delta f$]{\label{f:af}
      \includegraphics[width=0.48\textwidth]{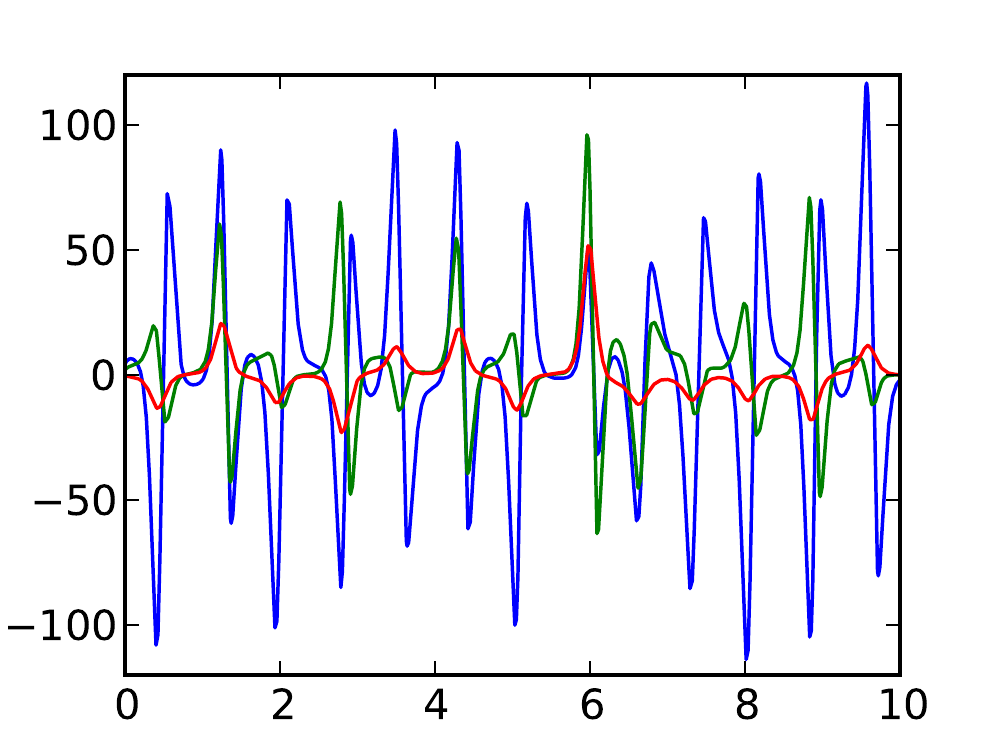}}
  \caption{Lyapunov vector decomposition of $\delta f$.
  The x-axes are $t$; the blue, green and red lines on the left are the
  first, second and third component of $\delta f$ as defined in Equation
  (\ref{lorenzdf});  the blue, green and red lines on
  the right are $a^f_1$, $a^f_2$ and $a^f_3$ in the decomposition
  of $\delta f$ (Equation (\ref{af})), respectively.}
\label{f:forwardaf}
\end{figure}

The forcing term defined in Equation (\ref{lorenzdf}) is plotted in
Figure \ref{f:df}.  Figure \ref{f:af} plots the decomposition
coefficients $a_i^f$, computed by solving a $3\times 3$ linear
system defined in Equation (\ref{af}) at every point on the trajectory.

\begin{figure}[htb]\centering
  \subfloat[$a^x_i,i=1,2,3$ for the $\delta f$]{\label{f:ax}
      \includegraphics[width=0.48\textwidth]{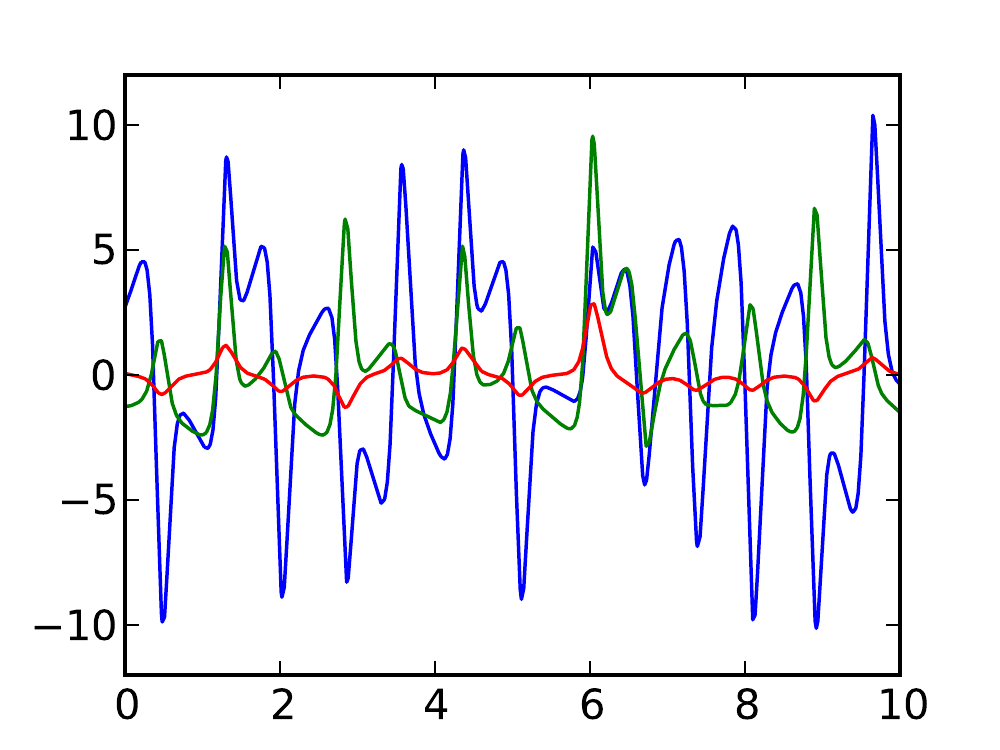}}
  \subfloat[$\delta x = \sum_{i=1}^3 a^x_i\, \phi_i$]{\label{f:dx}
      \includegraphics[width=0.48\textwidth]{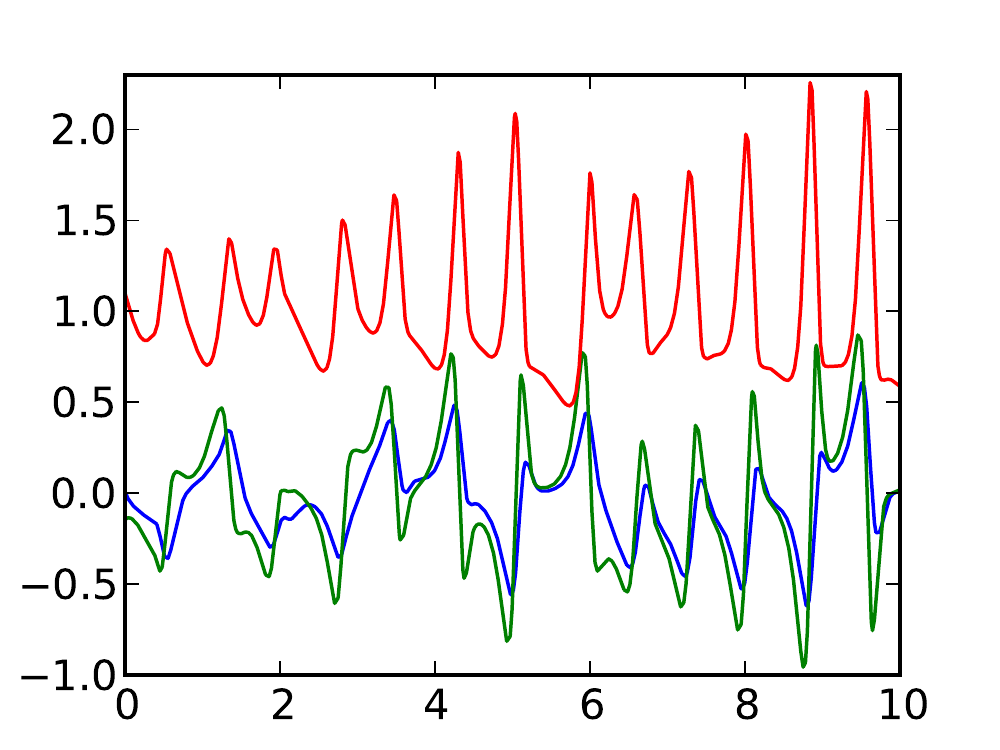}}
  \caption{Inversion of $S_f$ for $\delta x = S_f^{-1} \delta f$.
  The x-axes are $t$; the blue, green and red lines on the left
  are $a^x_1$, $a^x_2$ and $a^x_3$, respectively;
  the blue, green and red lines on the right are the first, second and
  third component of $\delta x$, computed via Equation (\ref{ax}).  }
\label{f:forwardax}
\end{figure}

For each $a^f_i$ obtained through the decomposition,
Equation (\ref{Sfode}) or (\ref{Sfoden0}) is solved to obtain
$a^x_i$.  For $i=1$, Equation (\ref{Sfode}) is solved backwards in time
from $t=T_A+T_B$ to $t=0$.  For $i=n_0=2$, the time compression
constant is estimated to be $\eta\approx -2.78$, and Equation
(\ref{Sfoden0}) is integrated to obtain $a^x_2$.
For $i=3$, Equation (\ref{Sfode}) is solved forward
in time from $t=-T_B$ to $t=T_A$.

The resulting values of $a^x_i,
i=1,2,3$ are plotted in Figure \ref{f:ax}.  These values are then
substituted into Equation (\ref{ax}) to obtain $\delta x$, as plotted
in Figure \ref{f:dx}.  The ``shadow'' trajectory defined as 
$x' = x + \epsilon \delta x$ is also plotted in Figure \ref{f:shadow} as
the red lines, for an $\epsilon=1/3$.  This $\delta x = S_f^{-1} \delta
f$ is approximately the shadow coordinate perturbation ``induced'' by a
$1/3$ increase in the input parameter $r$, a.k.a. the Rayleigh number
in the Lorenz attractor.

The last step of the forward sensitivity analysis algorithm is
computing the sensitivity derivatives of the output statistical
quantities using Equation (\ref{sens}).  We found that using a windowed
time averaging \cite{lcoadj_jcp} yields more accurate sensitivities.
Here our estimates over the time interval
$[0,T_A]$ are
\begin{equation} \label{sensResult1}
\frac{d \langle x_1^2\rangle}{dr} \approx 2.64 \;,\quad
\frac{d \langle x_2^2\rangle}{dr} \approx 3.99 \;,\quad
\frac{d \langle x_3\rangle}{dr}   \approx 1.01
\end{equation}
These sensitivity values compare well to results obtained through finite
difference, as shown in Section \ref{f:fd}.

\subsection{Adjoint Sensitivity Analysis}

\label{s:lorenzadjoint}

We demonstrate our adjoint sensitivity analysis algorithm
by computing the sensitivity
derivatives of the statistical quantity $\langle x_3\rangle$
to small perturbations in the three system parameters $s$, $r$ and $b$
in the Lorenz attractor Equation (\ref{lorenz}).

\begin{figure}[htb]\centering
  \subfloat[$\dfrac{\partial J}{\partial x}$ for $J = x_3$]{\label{f:dJdx}
      \includegraphics[width=0.48\textwidth]{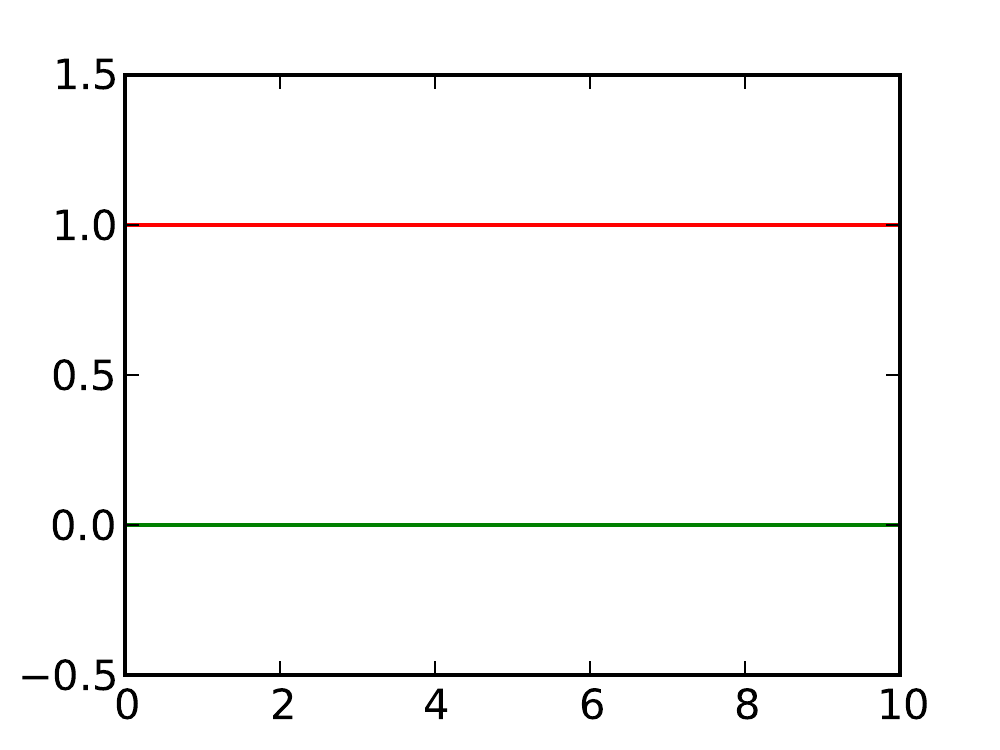}}
  \subfloat[$\hat{a}^x_i,i=1,2,3$ for the $\dfrac{\partial J}{\partial x}$]
  {\label{f:psiax}
      \includegraphics[width=0.48\textwidth]{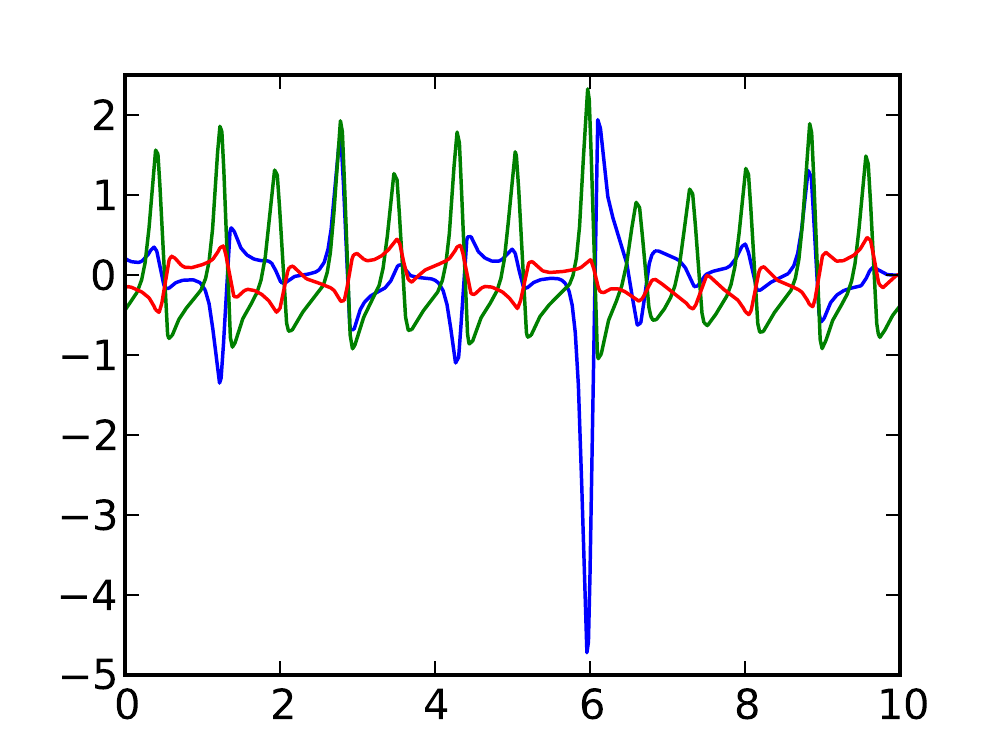}}
  \caption{Adjoint Lyapunov vector decomposition of $\partial J/\partial x$.
  The x-axes are $t$; the blue, green and red lines on the left are the
  first, second and third component of $\partial J/\partial x$;
  the blue, green and red lines on
  the right are $\hat{a}^x_1$, $\hat{a}^x_2$ and $\hat{a}^x_3$
  in the decomposition of $\partial J/\partial x$ (Equation (\ref{bx})),
  respectively.}
\label{f:adjointax}
\end{figure}

The first three steps of Algorithm \ref{alg:2} is the same as in
Algorithm \ref{alg:1}, and has been demonstrated in Section
\ref{s:decomp}.  Step \ref{alg2:step4} involves decomposing
$(\partial J/\partial x)^T$ into three adjoint Lyapunov covariant vectors.
In our case, $J(x) = x_3$, therefore $\partial J/\partial x \equiv (0,0,1)$,
as plotted in Figure \ref{f:dJdx}.  The adjoint Lyapunov covariant
vectors $\psi_i$ can be computed using Equation (\ref{conjugate}) by
inverting the $3\times 3$ matrix formed by the (primal) Lyapunov covariant
vectors $\phi_i$ at every point on the trajectory.
The coefficients $\hat{a}^x_i, i=1,2,3$ can then be computed by
solving Equation (\ref{bx}).
These scalar quantities along the trajectory are plotted
in Figure \ref{f:psiax} for $t\in [0,T_A]$.

\begin{figure}[htb]\centering
  \subfloat[$\hat{a}^f_i,i=1,2,3$ solved using Equation (\ref{afaxadj})]
  {\label{f:psiaf}
      \includegraphics[width=0.48\textwidth]{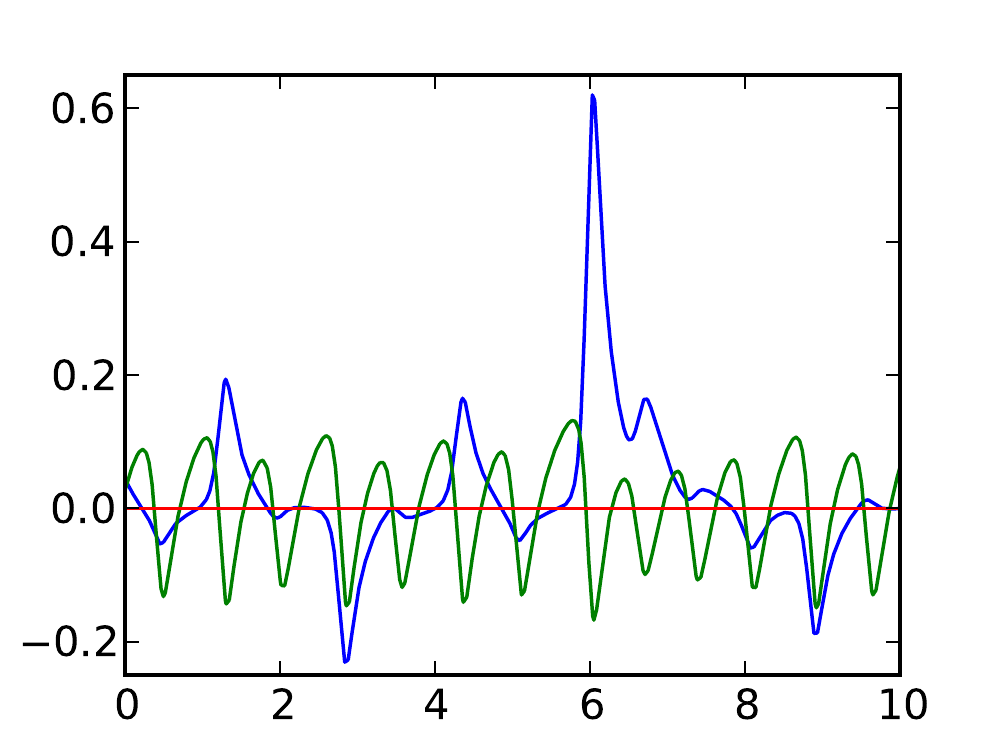}}
  \subfloat[$\hat{f} = \sum_{i=1}^3 \hat{a}^f_i\, \psi_i$]{\label{f:dJdf}
      \includegraphics[width=0.48\textwidth]{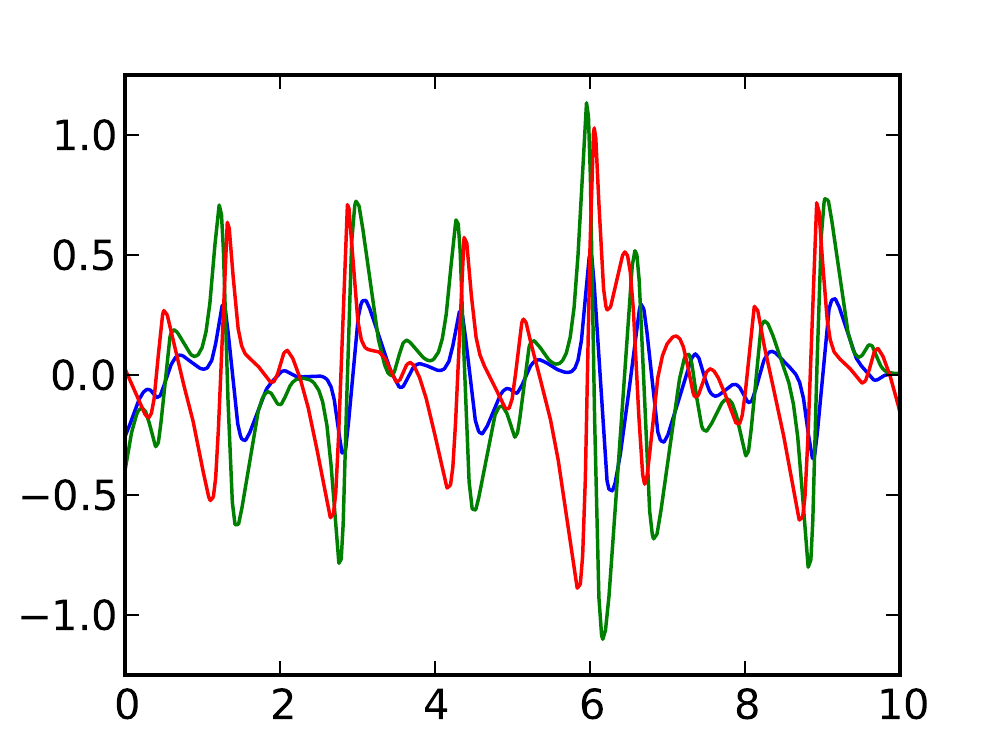}}
  \caption{Computation of the adjoint solution $\hat{f}$ for the Lorenz
  attractor.
  The x-axes are $t$; the blue, green and red lines on the left
  are $\hat{a}^f_1$, $\hat{a}^f_2$ and $\hat{a}^f_3$, respectively;
  the blue, green and red lines on the right are the first, second and
  third component of $\hat{f}$, computed via Equation (\ref{bf}).  }
\label{f:adjointaf}
\end{figure}

\begin{figure}[htb] \centering
\includegraphics[width=3.6in,trim=55 45 45 45,clip]{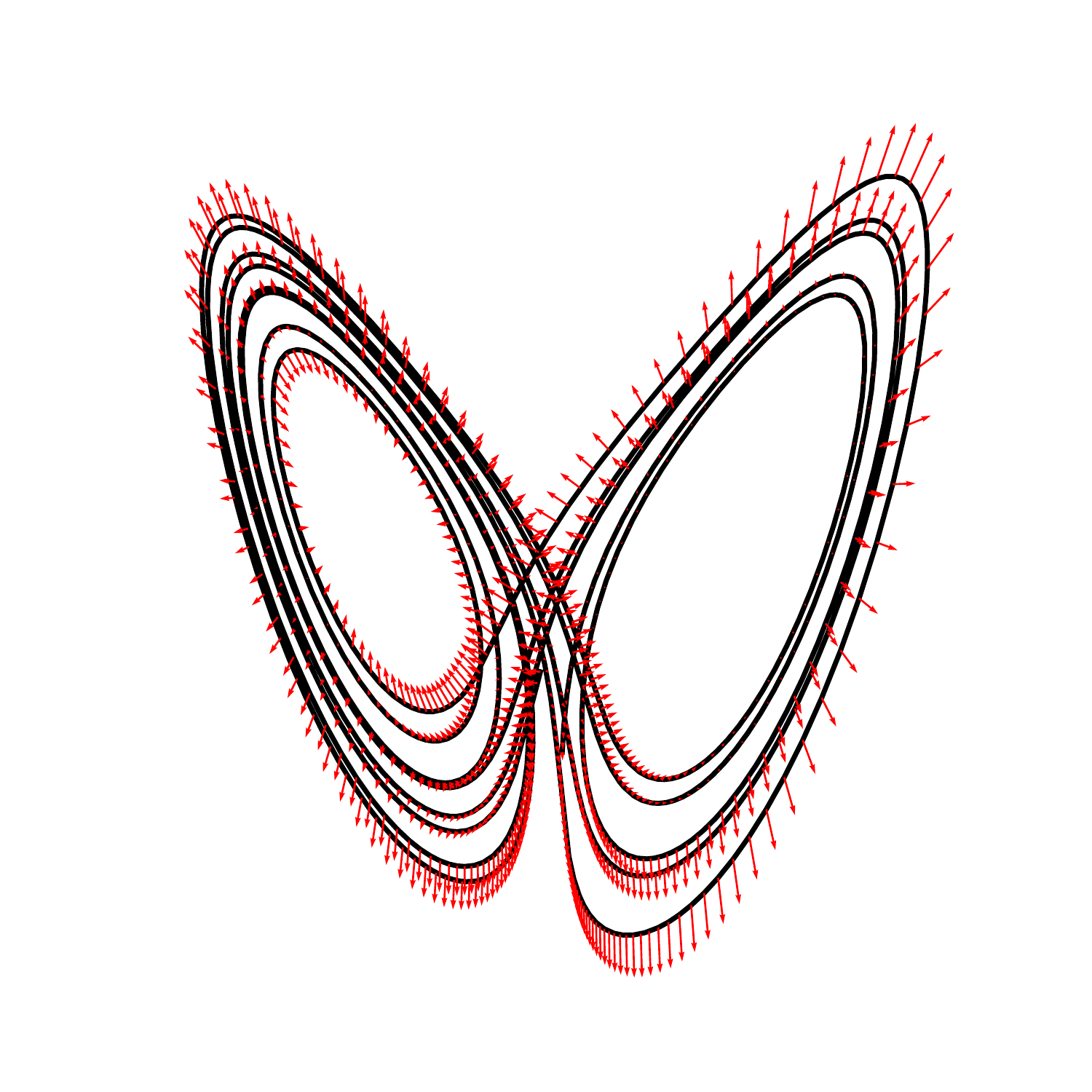}
\caption{The adjoint sensitivity derivative
$\hat{f}$ as in Equation (\ref{fadj}), represented by arrows on the
trajectory.}
\label{f:adjoint}
\end{figure}

Once we obtain $\hat{a}_i^x$, $\hat{a}_i^f$ can be computed by solving
Equation (\ref{afaxadj}).  The solution is plotted in Figure
\ref{f:psiaf}.  Equation (\ref{bf}) can then be used to combine
the $\hat{a}_i^f$ into the adjoint vector $\hat{f}$.
The computed $\hat{f}$ along the trajectory
is plotted both in Figure \ref{f:dJdf}
as a function of $t$, and also in Figure \ref{f:adjoint} as arrows
on the trajectory in the state space.

The last step of the adjoint sensitivity analysis algorithm is
computing the sensitivity derivatives of $\langle J\rangle$
to the perturbations $\delta f_s = \frac{df}{ds}$,
$\delta f_r = \frac{df}{dr}$ and $\delta f_b = \frac{df}{db}$
using Equation (\ref{fadj}).  Here our estimates over the time interval
$[0,T_A]$ are computed as
\begin{equation} \label{sensResult2}
\frac{d \langle x_3\rangle}{ds} \approx 0.21 \;,\quad
\frac{d \langle x_3\rangle}{dr} \approx 0.97 \;,\quad
\frac{d \langle x_3\rangle}{db} \approx -1.74
\end{equation}

Note that $\frac{d \langle x_3\rangle}{dr}$ estimated using adjoint
method differs from the same value estimated using forward method
(\ref{sensResult1}).
This discrepancy can be caused by the different numerical treatments to the
time dilation term in the two methods.
The forward method numerically estimates the time
dilation constant $\eta$ through Equation (\ref{eta});
while the adjoint method sets the mean of $\hat{a}_i^f$ to zero
(\ref{afaxadj0}), so that the computation is independent to the value of
$\eta$.  This difference 
could cause apparent discrepancy in the estimated sensitivity
derivatives.

The next section compares these sensitivity estimates, together with the
sensitivity estimates computed in Section \ref{s:lorenzforward},
to a finite difference study.

\subsection{Comparison with the finite difference method}

\label{f:fd}

To reduce the noise in the computed statistical quantities in
the finite difference study, a very long time integration length of
$T=100,000$ is used for each simulation.  Despite this long time
averaging, the quantities computed contain statistical noise of the
order $0.01$.  The noise limits the step size of the finite difference
sensitivity study.
Fortunately all the output statistical quantities seem fairly linear with
respect to the input parameters, and a moderately large step size
of the order $0.1$ can be used.
To further reduce the effect of statistical noise,
we perform linear regressions through
$10$ simulations of the Lorenz attractor, with $r$
equally spaced between $27.9$ and $28.1$.  The total time integration
length (excluding spin up time) is $1,000,000$.
The resulting computation
cost is in sharp contrast to our method, which involves a trajectory
of only length $20$.

Similar analysis is performed for the parameters $s$ and $b$, where
10 values of $s$ equally spaced between $9.8$ and $10.2$ are used,
and 10 values of $b$ equally spaced between $8/3-0.02$ and $8/3+0.02$
are used.  The slopes estimated from the linear
regressions, together with $3\sigma$ confidence intervals (where
$\sigma$ is the standard error of the linear regression) is listed below:
\begin{equation}\label{sensFD}
\begin{split}
& \frac{d \langle x_1^2\rangle}{dr} =  2.70 \pm 0.10  \;,\quad
  \frac{d \langle x_2^2\rangle}{dr} =  3.87 \pm 0.18  \;,\quad
  \frac{d \langle x_3\rangle}{dr}   =  1.01 \pm 0.04  \\
& \frac{d \langle x_3\rangle}{ds}   =  0.16 \pm 0.02  \;,\quad
  \frac{d \langle x_3\rangle}{db}   = -1.68 \pm 0.15  \;.
\end{split}
\end{equation}

\begin{figure}[htb]\centering
  \subfloat[$\dfrac{\partial \langle x_1^2\rangle}{\partial r}$]
  {\label{f:forxr}
      \includegraphics[width=0.33\textwidth]{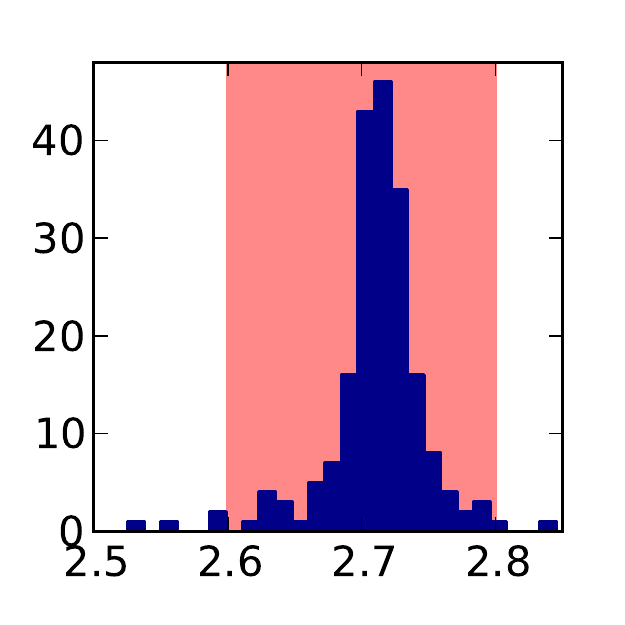}}
  \subfloat[$\dfrac{\partial \langle x_2^2\rangle}{\partial r}$]
  {\label{f:foryr}
      \includegraphics[width=0.33\textwidth]{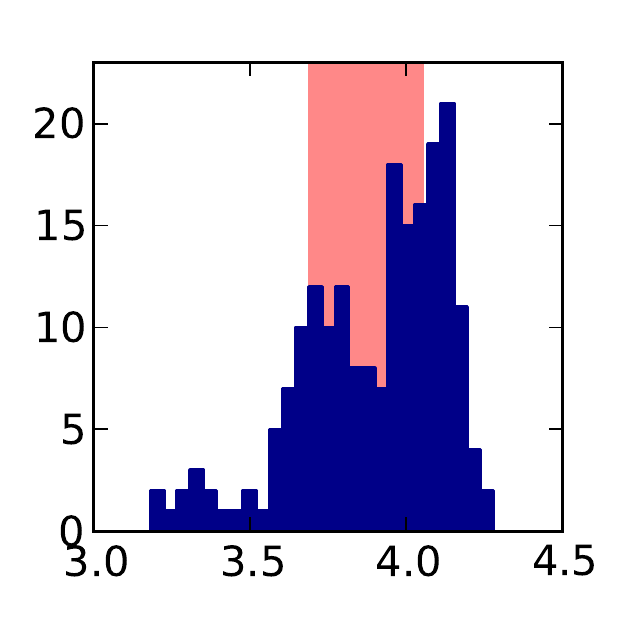}}
  \subfloat[$\dfrac{\partial \langle x_3\rangle}{\partial r}$]
  {\label{f:forzr}
      \includegraphics[width=0.33\textwidth]{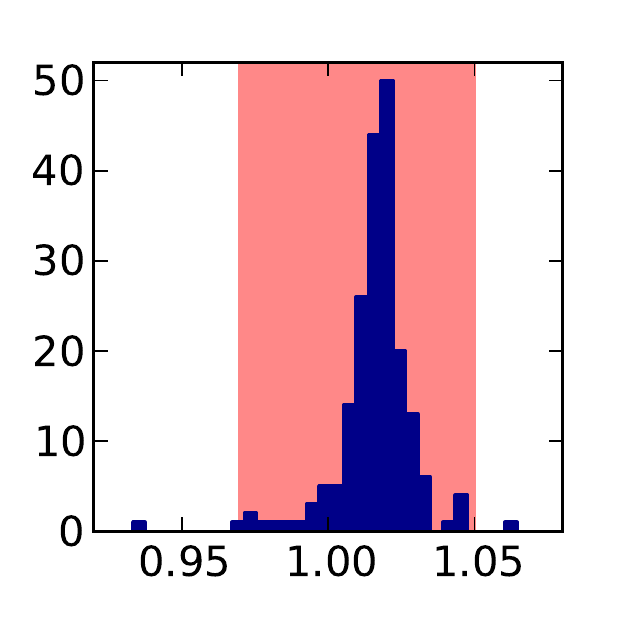}}
  \caption{Histogram of sensitivities computed using Algorithm \ref{alg:1}
  (forward sensitivity analysis) starting from 200 random initial conditions.
  $T_A=10, T_B=5$.
  The red region identifies the $3\sigma$ confidence
  interval estimated using finite difference regression.}
\label{f:stats}
\end{figure}

\begin{figure}[htb]\centering
  \subfloat[$\dfrac{\partial \langle x_3\rangle}{\partial s}$]{
  \label{f:adjzs}
      \includegraphics[width=0.33\textwidth]{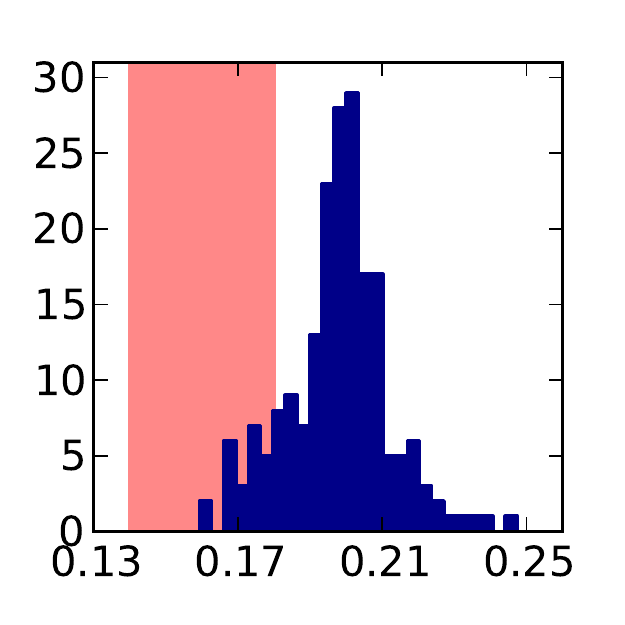}}
  \subfloat[$\dfrac{\partial \langle x_3\rangle}{\partial r}$]
  {\label{f:adjzr}
      \includegraphics[width=0.33\textwidth]{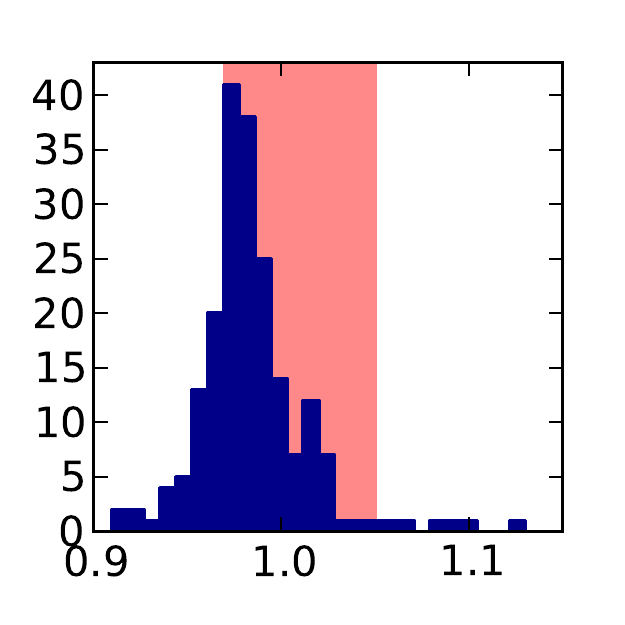}}
  \subfloat[$\dfrac{\partial \langle x_3\rangle}{\partial b}$]
  {\label{f:adjzb}
      \includegraphics[width=0.33\textwidth]{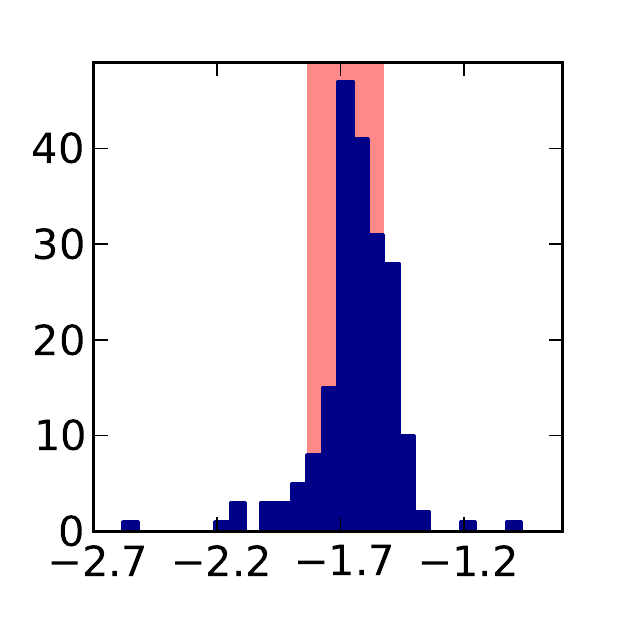}}
  \caption{Histogram of sensitivities computed using Algorithm \ref{alg:2}
  (adjoint sensitivity analysis) starting from 200 random initial conditions.
  $T_A=10, T_B=5$.
  The red region identifies the $3\sigma$ confidence
  interval estimated using finite difference regression.}
\label{f:statsadj}
\end{figure}

To further assess the accuracy of our algorithm, which involves finite
time approximations to Equations (\ref{sens}) and (\ref{fadj}),
we repeated both Algorithm \ref{alg:1} and Algorithm
\ref{alg:2} for 200 times, starting from random initial conditions at
$T=-10$.  We keep the statistical averaging time $T_A=10$ and
the spin up buffer time $T_B=5$.  The resulting histogram of sensitivities
computed with Algorithm \ref{alg:1} is shown in Figure \ref{f:stats};
the histogram of sensitivities computed with Algorithm \ref{alg:2}
is shown in Figure \ref{f:statsadj}.  The finite difference estimates
are also indicated in these plots.

We observe that our algorithms compute accurate sensitivities most of the
time.  However, some of the computed sensitivities seems to have heavy tails
in their distribution.
This may be due to behavior of the Lorenz attractor
near the unstable fixed point $(0,0,0)$.  Similar heavy tailed
distribution has been observed in other studies of the Lorenz attractor
\cite{eyinkclimate}.  They found that certain quantities computed on
Lorenz attractor can have unbounded second moment.  This could be the
case in our sensitivity estimates.  Despite this minor drawback,
the sensitivities computed using our algorithm have good quality.
Our algorithms are much more efficient
than existing sensitivity computation methods using ensemble averages.

\section{Conclusion}

This paper derived a forward algorithm and an adjoint algorithm
for computing sensitivity derivatives in chaotic dynamical systems.
Both algorithms efficiently compute the derivative of statistical
quantities $\langle J\rangle$ to infinitesimal perturbations
$\epsilon\,\delta f$ to the dynamics.

The forward algorithm starts from a
given perturbation $\delta f$, and computes a perturbed ``shadow''
coordinate system $\delta x$, e.g. as shown in Figure \ref{f:shadow}.
The sensitivity derivatives of multiple statistical quantities
to the given $\delta f$ can be computed from $\delta x$.
The adjoint algorithm starts from a statistical quantity $\langle
J\rangle$, and computes an adjoint vector $\hat{f}$, e.g. as
shown in Figure \ref{f:adjoint}.  The sensitivity derivative of
the given $\langle
J\rangle$ to multiple input perturbations can be computed
from $\hat{f}$.

We demonstrated both the forward and adjoint algorithms on the Lorenz
attractor at standard parameter values.  The forward sensitivity
analysis algorithm is used to simultaneously compute
$\frac{\partial\langle x_1^2\rangle}{\partial r}$,
$\frac{\partial\langle x_2^2\rangle}{\partial r}$,
and $\frac{\partial\langle x_3\rangle}{\partial r}$;
the adjoint sensitivity analysis algorithm is used to simultaneously
compute 
$\frac{\partial\langle x_3\rangle}{\partial s}$,
$\frac{\partial\langle x_3\rangle}{\partial r}$,
and $\frac{\partial\langle x_3\rangle}{\partial b}$.
We show that using a single trajectory of length about $20$,
both algorithms can efficiently compute accurate estimates of all the
sensitivity derivatives.

\label{s:conclusion}

\bibliography{master}

\appendix
\end{document}